\documentstyle[aps,epsf,amssymb]{revtex}
\begin{document}
\twocolumn
\title{Quantum cryptographic ranging}
\author{Vittorio Giovannetti$^1$, Seth Lloyd$^{1,2}$, and Lorenzo
Maccone$^1$}
\address{Massachusetts Institute of Technology,\\ $^1$Research
Laboratory of Electronics\\ $^2$Department of
Mechanical Engineering\\ Cambridge, MA 02139, USA.}
\maketitle
\begin{abstract}
{\bf Abstract.} We present a system to measure the distance between
two parties that allows only trusted people to access the result. The
security of the protocol is guaranteed by the complementarity
principle in quantum mechanics. The protocol can be realized with
available technology, at least as a proof of principle experiment.
\end{abstract}
\bigskip
\narrowtext
The following touching problem will be addressed in this paper. Alice
is lost in the woods. Her friend Bob needs to find her, rescue her,
and live happily ever after. On the other hand, the bad wolf Eve also
wants to find her, in order to gobble her up. Suppose for simplicity
an unidimensional forest, Alice and Bob need to find their relative
position (ranging) without giving any hint to Eve. This intent is
analogous to the one underlying cryptography, {i.e.} the exchange
of information in a secure fashion. In this respect we will refer to
it as crypto-positioning procedure.

In this paper, we show how the quantum mechanical time--energy
uncertainty principle can be employed to measure in a quantum
cryptographically secure fashion the distance between Alice and
Bob. This means that the physical limitations imposed by quantum
mechanics do not allow, not even in principle in the ideal case, any
kind of eavesdropping on Eve's part. The secure exchange of
information, based on analogous ``weird'' quantum effects has been
shown long ago in {\cite{bb84,b92,ekert}} and technological
applications seem to be at hand {\cite{rarity}}. Our idea to help Bob
in his quest stems by joining Ekert's quantum cryptographic protocol
{\cite{ekert}} with the recently proposed quantum positioning protocol
{\cite{nature}}, that allows one to perform ranging using
frequency-entangled states.

In {\cite{nature,qps}} we have shown how, by using $N$
frequency-entangled photons, one can obtain an $1/\sqrt{N}$ accuracy
enhancement in finding out the distance between Alice and Bob over the
case in which the $N$ photons are unentangled. This is a {\it truly}
quantum effect that arises from the strong photon correlations between
photons originating from the entanglement. This same fact, however,
makes the loss of a single photon critical: as shown in {\cite{qps}},
when one of the entangled photons that travel from Bob to Alice is
lost, the remaining photons yield no information at all on the
distance between the two. Such an apparent drawback turns out to be
the key feature in devising the crypto-positioning protocol.

In what follows we can limit our analysis to the case $N=2$. In this
situation it is possible to use the state generated by cw-pumped
spontaneous parametric down-conversion, in which the two generated
photons are anticorrelated in frequency, {i.e.}
\begin{eqnarray} |\Psi\rangle\equiv\int
d\omega\;\phi(\omega)|\omega_0+\omega\rangle_I
|\omega_0-\omega\rangle_S
\;\label{tbst}.
\end{eqnarray}
In Eq. (\ref{tbst}) the notation $|\nu\rangle$ refers to a single
photon state of frequency $\nu$, the ket subscripts refer to the two
distinct field modes (the signal $S$ and the idler $I$) generated by
the crystal, $2\omega_0$ is the pump frequency, and $\phi(\omega)$ is
the two-photon spectral function centered in $\omega=0$ with bandwidth
$\Delta\omega$. The state $|\Psi\rangle$ is one of the best known
sources of entanglement currently available and an enormous amount of
literature both theoretical and experimental is accessible (see for
example {\cite{mandel}} and citations therein). Another possibility
that can be exploited is the recently proposed `difference beam state'
{\cite{db}} that displays frequency {\it correlated} photons, whereas
$|\Psi\rangle$ displays anti-correlation in frequency.

The properties of the entanglement in the state $|\Psi\rangle$ are
such that, if one measures the frequency of the signal photon, he/she
would obtain a random value $\omega_0+\omega$ with probability density
$|\phi(\omega)|^2$, but the subsequent measurement on the idler photon
will have the predictable outcome $\omega_0-\omega$ (and vice versa
if the measurements are reversed). On the other hand, it is possible
to show that if one measures the time of arrival of the first photon
on a detector, then he/she will be able to predict (with an accuracy
of the order of $\Delta\omega^{-1}$) the time of arrival of the second
photon on a second detector at a distance $L$. In fact, the joint
probability of measuring the first photon at time $t_1$ and the second
at time $t_2$ is given by
\begin{eqnarray}
P_c(t_1,t_2)\propto\left|\int d\omega
\;\phi(\omega)\;e^{-i\omega(t_1-t_2+L/c)}
\right|^2
\;\label{pdic},
\end{eqnarray}
which exhibits a peak centered in $t_2-t_1=L/c$ of width proportional
to $\Delta\omega^{-1}$.  What happens when one measures the frequency
of the first photon and the time of arrival of the second? In this
case it is possible to show that the outcome of the time of arrival
measurement is completely unpredictable: all the timing information
has been ``erased'' by the frequency measurement. In this respect, the
measurement of the frequency on one of the photons has the same effect
as the loss of such photon.

\begin{figure}[hbt]
\begin{center}\epsfxsize=.6
\hsize\leavevmode\epsffile{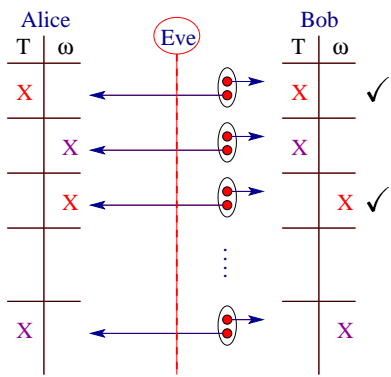} 
\end{center}
\caption{ Alice and Bob randomly choose to measure either the time of arrival
$T$ or the frequency $\omega$ on each copy of the two-photon state
$|\Psi\rangle$ they share. They retain only the copies for which their
choices agree, {i.e.} the checked (\checkmark) copies. }
\label{f:schema}\end{figure}

The crypto-positioning procedure, depicted in Fig. {\ref{f:schema}},
is the following:\begin{enumerate}\item Bob produces a certain number
of labeled copies of the two-photon state $|\Psi\rangle$. Of each copy
he sends Alice one of the two photons ({e.g.} the signal
photon).\item Of the idler photon he did not send Alice, he randomly
measures either the frequency or the time at which it reaches a
photodetector placed at a known distance from him.\item In the thick
of the woods, Alice receives Bob's photon and she also randomly
chooses to measure either the frequency or the time of arrival.\item
Alice and Bob broadcast the kind of measurement (frequency or time of
arrival) they performed on each of the two-photons copies. They
discard all the measurement results of the cases in which their
choices did not match.\item Alice and Bob exchange the results of the
frequency measurements and compare them. If the communication channel
is perfect and there is no eavesdropper measuring photon transit
times, these results are correlated (namely, if Alice had obtained the
frequency $\omega_0+\omega$, Bob must have found
$\omega_0-\omega$). On the contrary, if Eve measured the transit times
of the photons, she unavoidably spoiled the frequency
entanglement. Alice and Bob find this out when they compare their
data, since it will no longer be correlated.\item Once they have
verified that Eve is not tapping on the exchanged photons, Alice
broadcasts to Bob all the measurement results of the photon times of
arrival. This information is useless to anybody except Bob (who knows
the timing information of the other photon of each couple) and it
allows him to find the lost Alice through Eq. ({\ref{pdic}}) which
yields the distance $L$, given the photon times of arrival $t_1$ and
$t_2$.
\end{enumerate}

Notice that Eve can measure the frequency of the traveling photons
without being detected. This, however, does not allow her to gain any
information on the distance $L$. She will only succeed in disturbing
Alice and Bob's communication. 

The scheme can be straightforwardly adapted to much more complicated
scenarios. For example, one may tailor the entanglement to situations
in which multiple rescuers are present and they can obtain Alice's
position only if they meet and exchange their data in the spirit of
quantum secret sharing protocols. A discussion of the quantum
crypto-positioning protocol can be also found in {\cite{qps}}.

In conclusion, we have presented a protocol that, using the frequency
entangled state at the output of a parametric downconversion crystal,
allows one to perform quantum crypto-positioning. No matter how many
resources the evil Eve devotes to eavesdropping, she will not be able
to prevent a happy ending, since only Bob will find Alice!

This work was funded by the ARDA, NRO, and by ARO under a MURI
program.


\begin{references} 
\bibitem{bb84} C. H. Bennett, G. Brassard, and N. D. Mermin,
Phys. Rev. Lett. {\bf 68}, 557 (1992).
\bibitem{b92} C. H. Bennett, Phys. Rev. Lett. 68, 3121 (1992). 
\bibitem{ekert} A. K. Ekert, Phys. Rev. Lett. {\bf 67}, 661 (1991);
A.K. Ekert, J.G. Rarity, P.R.  Tapster, and G.M. Palma,
Phys. Rev. Lett. 69, 1293 (1992).
\bibitem{rarity} D. S. Naik, C. G. Peterson, A. G. White,
A. J. Berglund, and P. G. Kwiat, Phys. Rev. Lett. {\bf 84}, 4733
(2000); T. Jennewein, C. Simon, G. Weihs, H. Weinfurter, and A.
Zeilinger, Phys. Rev. Lett. {\bf 84}, 4729 (2000); G. Ribordy,
J. Brendel, J.-D. Gautier, N. Gisin, and H. Zbinden, Phys. Rev. A {\bf
63}, 012309 (2001).
\bibitem{nature} V. Giovannetti, S. Lloyd, and L. Maccone, Nature {\bf
412}, 417 (2001).
\bibitem{qps} V. Giovannetti, S. Lloyd, and L. Maccone, ``Positioning
and clock synchronization through entanglement'', Phys. Rev. A, to be
published.
\bibitem{mandel} L. Mandel  and  E. Wolf {\em Optical coherence and quantum 
optics} (Cambridge Univ. press, Cambridge, 1995).
\bibitem{db} V. Giovannetti, L. Maccone, J.  H. Shapiro,
and F. N. C. Wong, ``Generating Biphotons with Coincident Frequencies
via Parametric Downconversion'', Eprint quant-ph/0109135
\end{references}
\end{document}